\documentclass[showpacs,twocolumn]{revtex4}

\usepackage[usenames]{color}
\usepackage{bm}
\usepackage{amsmath}
\usepackage{amssymb}
\usepackage{epsfig}
\usepackage{graphics}

\newcommand{\bee}{\begin{eqnarray}}
\newcommand{\ene}{\end{eqnarray}}
\newcommand{\eqb}{\begin{equation}}
\newcommand{\eqe}{\end{equation}}

\begin{document}

\title{Quantum correlations and non-classicality in
a system of two coupled vertical external cavity surface emitting
lasers}

\author{D.~Mogilevtsev$^{1}$, D.~B.~Horoshko$^{1}$, Yu.~M.~Golubev$^{2}$, and M.~I.~Kolobov$^{3}$}

\affiliation{$^{1}$Institute of Physics, Belarus National Academy
of Sciences, F.Skarina Ave. 68, Minsk 220072 Belarus; \\
$^{2}$  St. Petersburg State University, 198504
Petrodvorets, St. Petersburg, Russia; \\
$^{3}$ Laboratoire PhLAM, Universit\'{e} Lille 1, F-59655
Villeneuve d'Ascq Cedex, France}

\begin{abstract}
Here we demonstrate that photocurrent noise reduction below the standard quantum limit and modal
anticorrelation can arise in two mode coupled two-VECSEL system
with common pump. This effect occurs due to correlated loss of laser modes. It is possible to suppress noise below the standard quantum limit even for Poissonian coherent
pumping, whereas the regularity of the pump can be harmful for
non-classicality.

\end{abstract}
 \maketitle

\section{Introduction}

Semiconductor lasers are by far the commonest lasers that
one finds around nowadays. However, despite being such a common
device, they are still actively researched, modified and improved.
They can be used for quantum
communication/informatics, in particular, for generating
non-classical states of radiation. Since the pioneering prediction
by Golubev and Sokolov \cite{golubev1984} (and it was a quickly
confirmed prediction  \cite{teich1985}), it is well known
that a semiconductor laser driven by a regular low-noise current is
able to produce photon-number squeezed states of light. One should have equal
amounts of emitters of the active media pumped in equal intervals
of time, so the emitted photons tend to be antibunched.

Here we demonstrate that the photocurrent noise suppression below the standard quantum limit
(which is usually associated with photon-number squeezing) in
semiconductor lasers (namely, in VECSEL lasers) can be reached by
an entirely different mechanism. Namely, it can occur due to
coupling of two modes to the same emitter with quickly decaying
populations and polarization (which we shall term here as
"correlated loss"). And for this kind of noise reduction to appear,
the regularity of the pump might be practically irrelevant.

It is well-known that coupling to the same emitter induces correlations between field reservoirs \cite{scully0}.
Also, correlated losses are quite common in situations, where one has two
or more systems (in our case, field modes) coupled to the third
lossy system. Interference arising in
this case  was shown to lead to entanglement between modes even in
absence of direct interaction between them
\cite{{entangle1},{entangle2}}. Correlated loss can lead to
appearance of nonlinear coupling between modes and even to
nonlinear loss producing nearly ideal Fock states
\cite{{mogilevtsev kerr 2009},{mogilevtsev opt lett 2010}}.
Notice, that to have a correlated loss with all consequent
nonlinear effects arising, one generally needs to have an hierarchy
of time-scales present in the system. Dynamics of the modes should
occur on much slower time-scale than dynamics of the
dissipating systems coupled to these modes.

Here we demonstrate that in the system of two coupled vertical
external cavity surface emitting lasers (VECSELs) with a common
pump, a correlated loss can arise and lead to appearance of
photocurrent noise reduction below the standard quantum limit (SQL). It occurs when this systems acts as a
class-A laser, and the population inversion lifetime is much
shorter, than the photon lifetimes in both modes. Since lasing emitters of the active media are coupled simultaneously
to both modes, quick decaying emitters disentangle from them giving rise to
effective non-linear coupling between modes (similarly to as
dispersive atom-field produces Kerr nonlinearity in EIT media
\cite{{eit-kerr1},{eit-kerr2},{eit-kerr3}}). Excited emitters of
the active media emit either in one or other mode without
possibility to re-absorb emitted photon. Such an interference of
emission channels gives rise to strong anti-correlation between
modes.  It was already demonstrated that VECSELs can be class-A
lasers \cite{{a-class1},{a-class2}}. Moreover, a scheme with
coupled VECSELs generating two output linearly polarized modes of
slightly different frequencies was recently realized in experiment
and demonstrated as class-A laser \cite{pal2010}. This scheme we
adopt as a basis for our theoretical model. We predict that in the
set-up similar to the one used in Ref. \cite{pal2010}, noise reduction below SQL might occur for each individual mode.
Also anticorrelation between modes can arise even for the Poissonian
pumping (for example, with the active region excited by an
external laser beam). However, our results are not limited to this
particular scheme, and can be easily generalized to any kin of lasing devices with correlated loss.

The outline of the paper is as follows. In the second Section we
introduce the model of two coupled VECSEL systems, write down
quantum Langevin equations for them, discuss  parameters of the
scheme and describe how the correlated loss and nonlinearities can
arise with the example of the simplified particular case of the
scheme. In the third Section we derive equations for collective
variables (polarizations and populations) and discuss a way to
describe pump statistics taking into account partition noise. In
Section IV we consider quasiclassical equations and demonstrate,
that equations for modal amplitudes can be reduced to equations
for A-class lasers. In Section V we investigate statistics of
small fluctuations around stationary values of modal amplitudes,
collective populations and polarizations, and discuss the obtained
results.

\section{The model of two coupled VECSELS}

\begin{figure}
\epsfig{ figure=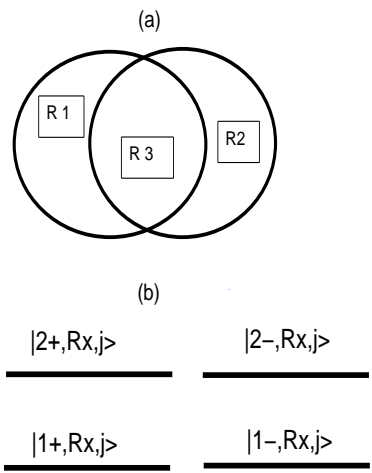,width=\linewidth,height=8cm} \caption
{(a) An illustration of the active media regions generating lasing
modes. Emitters of the region 1 (denoted as R1)  interact only
with modes denoted as "a", emitters of the region 2 (denoted as
R2) interact only with modes denoted as "b"; emitters of the
region 3 (denoted as R3) interact with both groups of modes. (b) A
scheme of levels of the $j$th emitter of region R$x$. \label{fig1} }
\end{figure}

Now let us consider a quantum model of two-frequency VECSEL in the
configuration described in the  paper \cite{pal2010}. There two
coupled VECSELs were created by using two-modal external cavity
with spatially separated modes. Spatial separation of modes was
achieved by introducing a birefringent crystal inside the laser
cavity. Modes overlap on the surface of the active media (which is
rather thin for VECSEL lasers), and the pumped region encompasses
all the surface. A schematic arrangement of the active media
regions participating in the generation process can be seen in
Fig.\ref{fig1}(a).

\subsection{Basic equations}

Considering our VECSEL model,
we use already well established four-level
spin-flip model for the description of emitters of the active media \cite{moloney1995}.
For the single-frequency
VECSEL the quantum theory 
was already extensively developed and elaborated on the basis of Langevin equation formalism
\cite{{kolobov2002},{kolobov2004}}. The model was shown to give
quite accurate description of VECSELs' lasing and
to represent adequately a complicated polarization dynamics appearing
in this kind of lasers. Each individual emitter/charge carrier of
the active semiconductor in this model is described by a
four-level system (actually, two two-level subsystems coupled by
spin-flip interaction). Lower levels of this model correspond to
the unexcited states of the semiconductor medium, i.e. without
electron-hole pairs. Upper levels correspond to the excited states
with electron-hole pairs. Two two-level subsystems are taken to
represent states with different (and opposite) angular momenta.
We assume that each two-level subsystem is coupled to the single
circularly polarized mode with a direction of polarization
corresponding to the angular momentum of the state, i.e. if one of
the two-level subsystems is coupled to the right-polarized mode,
than the other subsystem is coupled to the left-polarized
one. The scheme of levels of the medium is depicted in
Fig.\ref{fig1}(b).

Now we proceed writing down a system of quantum Langevin equations
for the interaction of the field with individual emitters along
the lines indicated in Ref.\cite{kolobov2002}. Let us introduce
bosonic annihilation operators, $\hat{a}_{\pm}$, $\hat{b}_{\pm}$
corresponding to  circularly polarized (in opposite directions for $+$ and $-$) modes of groups
"a" and "b". Then, for these modal operators we have the following system of equations \cite{kolobov2002}:
\begin{eqnarray}
\nonumber \frac{d\hat{a}_{\pm}}{dt}= -
(\kappa_a+i\omega_a)\hat{a}_{\pm}
-(\kappa_{ap}+i\omega_{ap}) \hat{a}_{\mp} - \\
\nonumber ig_a\Bigl(\sum\limits_{j\in R1}
\Theta(t-t^{(1)}_j)\sigma_{j\pm}^-+
\sum\limits_{l\in R3}
\Theta(t-t^{(3)}_l)\sigma_{l\pm}^-\Bigr) + \hat{f}_{a\pm}, \\
\frac{d\hat{b}_{\pm}}{dt}= - (\kappa_b+i\omega_b)\hat{b}_{\pm}
+(\kappa_{bp}+i\omega_{bp}) \hat{b}_{\mp} - \\
\nonumber ig_b\Bigl(\sum\limits_{k\in R2}\Theta(t-t^{(2)}_k)
\sigma_{k\pm}^- +
\sum\limits_{l\in R3}\Theta(t-t^{(3)}_l)
\sigma_{l\pm}^-\Bigr)+\hat{f}_{b\pm}, \label{field single}
\end{eqnarray}
where operators $\sigma_{j\pm}^-$ denote transition operators
from the upper to lower level of the $j$th emitter for corresponding circular polarizations. Using notations
of the state vectors corresponding to the
emitter levels given in Fig.\ref{fig1}, one writes
\[\sigma_{j\pm}^-=|1\pm,Rx,j\rangle\langle 2\pm,Rx,j|, \quad j\in Rx,\]
with $\sigma_{j\pm}^+$ being its Hermitian conjugate.
Quantities $\kappa_{a,b}$ are decay rates for modes $a,b$. We
assume that the linear dichroism is present, so, modes of
different linear polarization decay with different rates. This
difference is represented by parameters $\kappa_{ap}$ and
$\kappa_{bp}$.  Quantities $g_{a,b}$ are the emitter-field
interaction constants for corresponding modes with frequencies $\omega_{a}$ and $\omega_{b}$, respectively. Parameters $\omega_{ap}$ and
$\omega_{bp}$ represent coupling between modes of different circular
polarization appearing due to linear birefringence.

For the model we adopt the simple injection-type pumping: an
excited emitter (electron in the active zone and the hole in the
valent zone) appears at a random time-moment. Step-functions
$\Theta(t-t^{(x)}_k)$ are describing such a process of driving the
active media. So, the function $\Theta(t-t^{(x)}_k)$ describes
switching on the interaction of the field with the $k$th emitter
of the $x$th region. Statistics of time moments, $t^{(x)}_k$,
defines the type of the pumping (regular, Poissonian
or else; for the details see Ref.\cite{scully}).

Operators $\hat{f}_{a,b\pm}$ represent quantum Langevin forces
introduced into equations to account  for quantum noises
(in particular, to preserve correct commutation relation for modal
operators). They are introduced in the standard manner (see, for
example, a brilliant review by Luiz Davidovich \cite{davidovich}).
We shall specify their properties later.

Now let us proceed with equations for single-emitter transition
operators for different active regions depicted in Fig.\ref{fig1}
:
\begin{eqnarray}
\nonumber
\frac{d}{dt}\sigma_{j\pm}^-=-(\gamma_{\bot}+iw)\sigma_{j\pm}^-+\hat{f}_{j\pm}^{\sigma}+\\
\nonumber
ig_a\Theta(t-t^{(1)}_j)(\hat{n}_{j\pm}^{(2)}-\hat{n}_{j\pm}^{(1)})\hat{a}_{\pm}, \quad j\in R1,\\
\frac{d}{dt}\sigma_{j\pm}^-=-(\gamma_{\bot}+iw)\sigma_{j\pm}^-+\hat{f}_{j\pm}^{\sigma}+\\
\nonumber
ig_b\Theta(t-t^{(2)}_j)(\hat{n}_{j\pm}^{(2)}-\hat{n}_{j\pm}^{(1)})\hat{b}_{\pm}, \quad j\in R2\\
\nonumber
\frac{d}{dt}\sigma_{j\pm}^-=-(\gamma_{\bot}+iw)\sigma_{j\pm}^-+\hat{f}_{j\pm}^{\sigma}+\\
\nonumber
i\Theta(t-t^{(3)}_j)(\hat{n}_{j\pm}^{(2)}-\hat{n}_{j\pm}^{(1)})(g_a\hat{a}_{\pm}+g_b\hat{b}_{\pm}), \quad j\in R3.
\label{polarization single}
\end{eqnarray}
Here the parameter $\gamma_{\bot}$ is the dephasing rate of
two-level subsystems of the emitter; $w$ is the transition
frequency. Operators $\hat{n}_{j\pm}^{(y)}$ describe
population of $y$th level of $j$th emitter of the $x$th region:
\begin{eqnarray}
\nonumber
\hat{n}_{j\pm}^{(2)}=|2\pm,Rx,j\rangle\langle 2\pm,Rx,j|,\\
\nonumber \hat{n}_{j\pm}^{(1)}=|1\pm,Rx,j\rangle\langle
1\pm,Rx,j|,
\end{eqnarray}
where $j\in Rx$.
Operators $\hat{f}_{j\pm}^{\sigma}$ represent the corresponding
Langevin forces; these operators are independent. We discuss
them in the next Subsection.

Finally, let us write down equations for single-emitter operators
describing populations. For upper levels these are
\begin{eqnarray}
\nonumber
\frac{d}{dt}\hat{n}_{j\pm}^{(2)}=-\gamma_2\hat{n}_{j\pm}^{(2)}-\gamma_c(\hat{n}_{j\pm}^{(2)}-\hat{n}_{j\mp}^{(2)})+\\
\nonumber
ig_a\Theta(t-t^{(1)}_j)(\hat{a}^{\dagger}_{\pm})\sigma_{j\pm}^--\mathrm{h.c.})+\hat{f}_{j\pm}^{(2)}, \quad j\in R1,
\\
\frac{d}{dt}\hat{n}_{j\pm}^{(2)}=-\gamma_2\hat{n}_{j\pm}^{(2)}-\gamma_c(\hat{n}_{j\pm}^{(2)}-\hat{n}_{j\mp}^{(2)})+
\\
\nonumber
ig_b\Theta(t-t^{(2)}_j)(\hat{b}^{\dagger}_{\pm})\sigma_{j\pm}^--\mathrm{h.c.})+\hat{f}_{j\pm}^{(2)}, \quad j\in R2,
\\
\nonumber
\frac{d}{dt}\hat{n}_{j\pm}^{(2)}=-\gamma_2\hat{n}_{j\pm}^{(2)}-\gamma_c(\hat{n}_{j\pm}^{(2)}-\hat{n}_{j\mp}^{(2)})+
\\
\nonumber
+i\Theta(t-t^{(3)}_j)((g_a\hat{a}^{\dagger}_{\pm}+g_b\hat{b}^{\dagger}_{\pm})\sigma_{j\pm}^--\mathrm{h.c.})
+\hat{f}_{j\pm}^{(2)}, \quad j\in R3. \label{upper population single}
\end{eqnarray}
Here $\gamma_2$ is the decay rate of the emitter's upper levels;
$\gamma_c$ is the rate of spin-flips between  levels $|2+,Rx,j\rangle$ and $|2-,Rx,j\rangle$. Operators $\hat{f}_{j\pm}^{(2)}$ represent corresponding Langevin
forces.

Equations for populations of lower levels are as follows:
\begin{eqnarray}
\nonumber
\frac{d}{dt}\hat{n}_{j\pm}^{(1)}=-\gamma_1\hat{n}_{j\pm}^{(1)}+\hat{f}_{j\pm}^{(1)}-\\
\nonumber
ig\Theta(t-t^{(1)}_j)(\hat{a}^{\dagger}_{\pm}\sigma_{j\pm}^--\mathrm{h.c.}), \quad j\in R1,
\\
\frac{d}{dt}\hat{n}_{j\pm}^{(1)}=-\gamma_1\hat{n}_{j\pm}^{(1)}+\hat{f}_{j\pm}^{(1)}-\\
\nonumber
ig\Theta(t-t^{(2)}_j)(\hat{b}^{\dagger}_{\pm}\sigma_{j\pm}^--\mathrm{h.c.}), \quad j\in R2,
\\
\nonumber
\frac{d}{dt}\hat{n}_{j\pm}^{(1)}=-\gamma_1\hat{n}_{j\pm}^{(1)}-+\hat{f}_{j\pm}^{(1)}-\\
\nonumber
ig\Theta(t-t^{(3)}_j)((\hat{a}^{\dagger}_{\pm}+\hat{b}^{\dagger}_{\pm})\sigma_{j\pm}^--\mathrm{h.c.}), \quad j\in R3.
\label{lower population single}
\end{eqnarray}
Here $\gamma_1$ is the decay rate of the emitter's lower levels;
operators $\hat{f}_{j\pm}^{(1)}$ represent corresponding Langevin forces.
Notice, that we are assuming no spin-flips between lower levels of
emitters. It is really unimportant for the considered scheme
because the decay rate of lower levels, $\gamma_1$, is taken to be
far exceeding that of upper levels, $\gamma_2$.

\subsection{Langevin forces}

Here we describe Langevin forces  in Eqs. (\ref{field
single},\ref{polarization single},\ref{upper population
single},\ref{lower population single}). First of all, they are
$\delta$-correlated, i.e. for any two forces $\hat{f}_x(t)$ and
$\hat{f}_y(\tau)$ one has
\[\langle
\hat{f}_x(t)\hat{f}_y(\tau)\rangle=d_{xy}(t)\delta(t-\tau),
\]
where brackets $\langle\ldots\rangle$ denote quantum averaging.
So, $d_{xt}(t)$ are c-number functions. However, they do generally
depend on stochastic variables, namely, emitter arrival times.
First-order averages of  Langevine forces are zero: $\langle
\hat{f}_x(t)\rangle=0$.

Then, we assume that  Langevin forces for different emitters of
different regions are independent quantum variables, so, for any
two forces corresponding to variables of different emitters,
$\hat{f}_{Rj}(t)$ and $\hat{f}_{Rk}(\tau)$, $j\neq k$, one has
$\langle\hat{f}_{Rj}(t)\hat{f}_{Rk}(\tau)\rangle=0$. Equally,
quantum Langevin forces for modes of $a$ and $b$ groups are
independent.

Thus, we proceed to non-zero second-order correlation
functions along the lines given in
Refs.\cite{{kolobov2002},{davidovich}}. Assuming that there is
practically no thermal noise of field modes at optical frequencies, one has
\begin{eqnarray}
\nonumber
\langle\hat{f}_{x\pm}(t)\hat{f}_{x\pm}^{\dagger}(\tau)\rangle=2\kappa_x\delta(t-\tau),
\\
\langle\hat{f}_{x\pm}(t)\hat{f}_{x\mp}^{\dagger}(\tau)\rangle=2\kappa_{xp}\delta(t-\tau),
\label{langevin field}
\end{eqnarray}
where $x=a,b$.

 The operators of Langevin forces for emitters' populations are
self-conjugated, so in all regions one has
\begin{eqnarray}
\nonumber
\langle\hat{f}_{j\pm}^{(2)}(t)\hat{f}_{j\pm}^{(2)}(\tau)\rangle=\delta(t-\tau)\times\\
\nonumber \left[\gamma_2\langle\hat{n}_{j\pm)}^{(2)}\rangle+
\gamma_c(\langle\hat{n}_{j\pm}^{(2)}\rangle+\langle\hat{n}_{j\mp}^{(2)}\rangle)\right],
\\
\langle\hat{f}_{j\pm}^{(2)}(t)\hat{f}_{j\mp}^{(2)}(\tau)\rangle=-\delta(t-\tau)\times\\
\nonumber
\gamma_c(\langle\hat{n}_{j\pm}^{(2)}\rangle+\langle\hat{n}_{j\mp}^{(2)}\rangle),\\
\nonumber
\langle\hat{f}_{j\pm}^{(1)}(t)\hat{f}_{j\pm}^{(1)}(\tau)\rangle=\gamma_1\langle\hat{n}_{j\pm}^{(1)}\rangle\delta(t-\tau).
\label{langevin population single}
\end{eqnarray}

For transition operators one has in the same way
\begin{eqnarray}
\nonumber
\Bigl\langle\left[\hat{f}_{j\pm}^{\sigma}(t)\right]^{\dagger}\hat{f}_{j\pm}^{\sigma}(\tau)\Bigr\rangle=\delta(t-\tau)\times\\
\nonumber
\left[(2\gamma_{\bot}-\gamma_2-\gamma_c)\langle\hat{n}_{j\pm}^{(2)}\rangle+
\gamma_c\langle\hat{n}_{j\mp}^{(2)}\rangle\right], \\
\Bigl\langle\hat{f}_{j\pm}^{\sigma}(t)\left[\hat{f}_{j\pm}^{\sigma}(\tau)\right]^{\dagger}\Bigr\rangle=\delta(t-\tau)\times\\
\nonumber
(2\gamma_{\bot}-\gamma_1)\langle\hat{n}_{j\pm}^{(1)}\rangle.
\label{langevin polarizations single}
\end{eqnarray}

Finally, we write down cross-correlations of Langevin forces for
transition operators and populations:
\begin{eqnarray}
\nonumber
\langle\hat{f}_{j\pm}^{\sigma}(t)\hat{f}_{j\pm}^{(2)}(\tau)\rangle=(\gamma_2+\gamma_c)
\langle\sigma_{j\pm}\rangle\delta(t-\tau)\\
\langle\hat{f}_{j\pm}^{\sigma}(t)\hat{f}_{j\mp}^{(2)}(\tau)\rangle=-\gamma_c
\langle\sigma_{j\pm}\rangle\delta(t-\tau),\\
\nonumber
\Bigl\langle\left[\hat{f}_{j\pm}^{\sigma}(t)\right]^{\dagger}\hat{f}_{j\pm}^{(1)}(\tau)\Bigr\rangle=
\gamma_1\langle\sigma_{j\pm}^{\dagger}\rangle\delta(t-\tau),
\label{langevin cross population polarization single}
\end{eqnarray}
also valid for all three regions.

\subsection{Parameters of the scheme}

We consider realistic values of parameters of the scheme outlined above in
this Section. They are close to those of the
experiment described in Ref. \cite{pal2010}. For these values the scheme acts as
a class-A laser. First of all, we assume that
the lower levels of emitters are emptied very rapidly, i.e.
$\gamma_1$ is sufficiently large to exclude
adiabatically the lower level populations. Actually, rapid decay of
lower level populations is a condition of this
model applicability for semiconductor lasers; see, for example,
Ref.\cite{jacobino1997}. Then, we make  a realistic
assumption about dephasing in the active media being very rapid,
so the decay rate, $\gamma_{\bot}$,  far exceeds the upper levels
decay rate, $\gamma_2$. Finally, we make an assumption of the
class-A laser: we take upper level decay rate to be far exceeding
modal decay rates, $\kappa_{a,b}$ and $\kappa_{ap,bp}$. So, the
following hierarchy for the time-scales takes place in our model:
\begin{equation}
\kappa_{a,b},\kappa_{ap,bp} \ll \gamma_2 \ll \gamma_{\bot},
\gamma_1. \label{hierarchy}
\end{equation}
Also, we assume that the interaction constants $g_{a,b}$ are
small in comparison with the the upper levels decay rate,
$\gamma_2$.

\subsection{Toy model for correlated loss}
\label{simple model}

Now let us demonstrate that satisfying the condition
(\ref{hierarchy}), one can get a nonlinear coupling between modes.
For illustration we take the simplified model of just two field
modes of the same frequency coupled resonantly to a single
two-level system subjected to strong dephasing and population
loss.
We describe this simplified model with the following master
equation for the density matrix, $\rho$, written in the basis
rotated with the emitters transition frequency
\begin{eqnarray}
\nonumber {d\over dt}
\rho=-ig[\sigma^+(a+b)+h.c.,\rho]+\\
(\gamma_{\bot}\mathcal{L}(\sigma_z)
+\gamma_{2}\mathcal{L}(\sigma^-))\rho,
 \label{model master equation1}
\end{eqnarray}
where the dissipator $\mathcal{L}(x)\rho=2x\rho
x^{\dagger}-x^{\dagger}x\rho-\rho x^{\dagger}x$.

The condition (\ref{hierarchy}) allows one to eliminate
adiabatically off-diagonal matrix elements $\langle k
|\rho|l\rangle$; where $k,l=1,2;\quad k\neq l$, since they decay
with a large rate dephasing rate, $\gamma_{\bot}$. For times
much exceeding $\gamma_{\bot}^{-1}$ one can write
\[\langle k|\rho|l\rangle\approx -
i\frac{g^2}{\gamma_{\bot}}\langle k|\hat{n}\sigma_z\rho|l\rangle,
\]
where $\hat{n}=(a^{\dagger}+b^{\dagger})(a+b)$.

So, the master equation is reduced to one describing a
dispersive coupling of modes with a two-level system:
\begin{eqnarray}
 {d\over dt} \rho \approx
-2i\frac{g^2}{\gamma_{\bot}}[\hat{n}\sigma_z,\rho]+
\gamma_2\mathcal{L}(\sigma^-)\rho \label{model master equation2}
\end{eqnarray}

Obviously, a coupling
between modes has arisen as a consequence of correlated loss. So,
modes become correlated even in the absence of direct coupling
between them as the result of strong decay of the emitters
coupled to the modes.

Moreover, under conditions leading to the dispersive coupling similar to
Eq.(\ref{model master equation2}), Kerr nonlinearities can
arise, too \cite{{eit-kerr1},{eit-kerr2},{eit-kerr3}}. Indeed,
adiabatically eliminating emitter variables from Eq. (\ref{model
master equation1}) while retaining terms up to
$g^3/\gamma_{\bot}^3$ and averaging over the state of emitter, it
is not hard to obtain for the reduced density matrix the following expression
\begin{eqnarray}
{d\over dt} \hat{\rho} \approx
-2i\frac{g^2}{\gamma_{\bot}}[\hat{n}, \hat{\rho}]-
4i\frac{g^4}{\gamma_{\bot}^3}[\hat{n}^2, \hat{\rho}].\label{model
master equation3}
\end{eqnarray}

Thus, one has both linear and cross-Kerr coupling between the modes.
Provided that modal losses are sufficiently low, self-Kerr and
cross-Kerr nonlinearities appearing in Eq.(\ref{model master
equation3}) are known to be able to lead both to photon number
squeezing of individual modes and to anticorrelations of modes
\cite{{eit-kerr1},{eit-kerr2},{eit-kerr3},{kerr_all}}. As it will
be seen below, it is just the case for our class-A coupled VECSELs
lasers.

\section{Collective equations}

The next step is to move from single-emitter equations to
equations for collective operators describing ensembles of
emitters in different regions of active media. Such collective
operators can be introduced in a standard manner (see, for
example, Refs.\cite{{kolobov2002},{scully},{davidovich}}). 
So, introducing for clarity different notations for different regions, we obtain for polarizations 
\begin{eqnarray}
\nonumber \hat{P}_{\pm}=-i\sum\limits_{j\in R1}
\Theta(t-t^{(1)}_j)\sigma_{j\pm}^-, \\
\hat{Q}_{\pm}=-i\sum\limits_{j\in R2}
\Theta(t-t^{(2)}_j)\sigma_{j\pm}^-,  \\
\nonumber \hat{\Xi}_{\pm}=-i\sum\limits_{j\in R3}
\Theta(t-t^{(3)}_j)\sigma_{j\pm}^-.
 \label{collective operators_p}
\end{eqnarray}
For collective population operators we have
\begin{eqnarray}
\nonumber \hat{M}_{y\pm}=\sum\limits_{j\in R1}
\Theta(t-t^{(1)}_j)\hat{n}_{j\pm}^{(y)}, \\
\hat{N}_{y\pm}=\sum\limits_{j\in R2}
\Theta(t-t^{(2)}_j)\hat{n}_{j\pm}^{(y)},  \\
\nonumber \hat{\Lambda}_{y\pm}=\sum\limits_{j\in R3}
\Theta(t-t^{(3)}_j)\hat{n}_{j\pm}^{(y)}.
 \label{collective operators_n}
\end{eqnarray}
where $y=1,2$ denotes the emitter level.

In this section we derive equations for collective variables
(\ref{collective operators_p},\ref{collective operators_n}), perform averaging over arrival times
of emitters via injection-like pumping and calculate quantum
Langevin forces corresponding to the introduced collective
operators.

\subsection{Equations for collective variables}

First of all, let us write equations for modal operators. From the
system (\ref{field single}) one has
\begin{eqnarray}
\nonumber \frac{d\hat{a}_{\pm}}{dt}= -
(\kappa_a+i\omega_a)\hat{a}_{\pm} +(\kappa_{ap}+i\omega_{ap})
\hat{a}_{\mp} + \\
\nonumber g_a(\hat{P}_{\pm}+
\hat{\Xi}_{\pm}) + \hat{f}_{a\pm}, \\
\frac{d\hat{b}_{\pm}}{dt}= - (\kappa_b+i\omega_b)\hat{b}_{\pm}
+(\kappa_{bp}+i\omega_{bp}) \hat{b}_{\mp} + \\
\nonumber g_b(\hat{Q}_{\pm} + \hat{\Xi}_{\pm})+\hat{f}_{b\pm},
\label{field collective}
\end{eqnarray}
As follows from Eqs.(\ref{polarization single},\ref{collective
operators_p},\ref{collective
operators_n}), equations for collective polarization operators are
\begin{eqnarray}
\nonumber
\frac{d}{dt}\hat{P}_{\pm}=-(\gamma_{\bot}+iw)\hat{P}_{\pm}+\\
\nonumber
g_a(\hat{M}_{2\pm}-\hat{M}_{1\pm})\hat{a}_{\pm}
+\hat{F}_{\pm}^{P}, \\
\frac{d}{dt}\hat{Q}_{\pm}=-(\gamma_{\bot}+iw)\hat{Q}_{\pm}+\\
\nonumber
g_b(\hat{N}_{2\pm}-\hat{N}_{1\pm})\hat{b}_{\pm}
+\hat{F}_{\pm}^{Q}, \\
\nonumber
\frac{d}{dt}\hat{\Xi}_{\pm}=-(\gamma_{\bot}+iw)\hat{\Xi}_{\pm}+\\
\nonumber
(\hat{\Lambda}_{2\pm}-\hat{\Lambda}_{1\pm})(g_a\hat{a}_{\pm}+g_b\hat{b}_{\pm})
+\hat{F}_{\pm}^{\Xi}. \label{polarization collective}
\end{eqnarray}
Collective Langevin operators for these equations are
\begin{eqnarray}
\nonumber \hat{F}_{\pm}^{P,Q,\Xi}=\sum\limits_{j\in
R1,R2,R3}\delta(t-t^{(1,2,3)}_j)\sigma_{j\pm}^-+\\
\sum\limits_{j\in
R1,R2,R3}\Theta(t-t^{(1,2,3)}_j)\hat{f}_{j\pm}^{\sigma}.
 \label{langevin polarization
collective}
\end{eqnarray}
Correlation properties for these operators will be given in the
next Subsection.

Equations for collective populations are not so trivial to derive
as those given above. Eqs. (\ref{field
collective},\ref{polarization collective}) remain invariant after
averaging over arrival times (of course, it holds, if one assumes
that operators in these equations are not correlated through it;
such an assumption is obviously valid near the stationary regime
that we are mostly interested in; see  Ref.\cite{scully}). It is not so for
equations describing collective populations, since the noise for
populations is biased (has non-zero average) due to presence of
the pump. We assume that emitters arrive fully excited, and become with
equal probability either $+$ or $-$ excited subsystem of
each VECSEL. Let us for the time being denote these biases for the
upper-level population collective noises as $R_x$, $x$, denoting
regions as in Fig.\ref{fig1}. As follows from Eqs.(\ref{upper
population single},\ref{collective
operators_p},\ref{collective
operators_n}), equations for
operators of total upper-state populations are
\begin{eqnarray}
\nonumber
\frac{d}{dt}\hat{M}_{2\pm}=R_1-\gamma_2\hat{M}_{2\pm}-\gamma_c(\hat{M}_{2\pm}-\hat{M}_{2\mp})-\\
\nonumber
g_a(\hat{a}_{\pm}^{\dagger}\hat{P}_{\pm}+\hat{P}^{\dagger}_{\pm}\hat{a}_{\pm})
+\hat{F}_{2\pm}^{M}, \\
\frac{d}{dt}\hat{N}_{2\pm}=R_2-\gamma_2\hat{N}_{2\pm}-\gamma_c(\hat{N}_{2\pm}-\hat{N}_{2\mp})-\\
\nonumber
g_b(\hat{b}_{\pm}^{\dagger}\hat{Q}_{\pm}+\hat{Q}^{\dagger}_{\pm}\hat{b}_{\pm})
+\hat{F}_{2\pm}^{N}, \\
\nonumber
\frac{d}{dt}\hat{\Lambda}_{2\pm}=R_3-\gamma_2\hat{\Lambda}_{2\pm}-\gamma_c(\hat{\Lambda}_{2\pm}-\hat{\Lambda}_{2\mp})-\\
\nonumber
((g_a\hat{a}_{\pm}^{\dagger}+g_b\hat{b}_{\pm}^{\dagger})\hat{\Xi}_{\pm}+\hat{\Xi}^{\dagger}_{\pm}(g_a\hat{a}_{\pm}+g_b\hat{b}_{\pm}))
+\hat{F}_{2\pm}^{\Lambda}. \label{upper level population collective}
\end{eqnarray}
In the system (\ref{upper level population collective}) the collective
Langevin operators are
\begin{eqnarray}
\nonumber \hat{F}_{2\pm}^{M,N,\Lambda}=\sum\limits_{j\in
R1,R2,R3}\delta(t-t^{(1,2,3)}_j)\hat{n}^{(2)}_{j\pm}+\\
\sum\limits_{j\in R1,R2,R3}\Theta(t-t^{(1,2,3)}_j)\hat{f}_{j\pm}^{(2)}
- R_{1,2,3}. \label{langevin upper state population collective}
\end{eqnarray}
From the equation  (\ref{langevin upper state population
collective}) it is clear that parameters $R_{1,2,3}$  are average
pump rates (i.e. emitter injection rates) of emitters of the certain
type (i.e. $+$ or $-$) in corresponding regions.
Indeed, it is easy to see that with our way of driving one has (see the derivation in Ref. \cite{kolobov2002})
\[\left\langle \sum\limits_{j\in R1,R2,R3}\delta(t-t^{(1,2,3)}_j)\hat{n}_{j\pm}^{(2)}\right\rangle_s=R_{1,2,3},
\]
where $\langle \ldots\rangle_s$ denoted averaging over the arrival
times.

In our scheme of two-frequency VECSEL  the pumping source for all three regions is
the same \cite{pal2010}. So, respective pump rates are not independent
parameters. Since we are assuming that density of emitters is the
same in all considered regions, the average pump rate for every
region is simply proportional to the size of this region. Thus, we
write
\begin{equation}
R_3=\sqrt{\xi_a}(R_1+R_3)=\sqrt{\xi_b}(R_2+R_3) \label{pump rates}
\end{equation}
where parameters $\xi_{a,b}$ describe the respective sizes of the
regions of modal overlap, $R_3$, relative to total sizes of
regions interacting with modes "a" and "b".

Finally, we write down equations for lower level collective
populations deriving them from Eqs.(\ref{lower population
single}). Notice that emitters are taken to be injected being
completely excited, i.e. they are on upper levels. Thus, we have
\begin{eqnarray}
\nonumber
\frac{d}{dt}\hat{M}_{1\pm}=-\gamma_1\hat{M}_{1\pm}+ \hat{F}_{1\pm}^{M}+\\
\nonumber
g_a(\hat{a}_{\pm}^{\dagger}\hat{P}_{\pm}+\hat{P}^{\dagger}_{\pm}\hat{a}_{\pm}), \\
\frac{d}{dt}\hat{N}_{1\pm}=-\gamma_1\hat{N}_{1\pm}+ \hat{F}_{1\pm}^{N}+\\
\nonumber
g_b(\hat{b}_{\pm}^{\dagger}\hat{Q}_{\pm}+\hat{Q}^{\dagger}_{\pm}\hat{b}_{\pm}), \\
\nonumber
\frac{d}{dt}\hat{\Lambda}_{\pm}=-\gamma_2\hat{\Lambda}_{\pm}+\hat{F}_{1\pm}^{\Lambda}+\\
\nonumber
((g_a\hat{a}_{\pm}^{\dagger}+g_b\hat{b}_{\pm}^{\dagger})\hat{\Xi}_{\pm}+\hat{\Xi}^{\dagger}_{\pm}(g_a\hat{a}_{\pm}+g_b\hat{b}_{\pm})).
\label{lower level population collective}
\end{eqnarray}
In this system the collective Langevin operators are
\begin{eqnarray}
\nonumber
 \hat{F}_{1\pm}^{M,N,\Lambda}=\sum\limits_{j\in
R1,R2,R3}\delta(t-t^{(1,2,3)}_j)\hat{n}_{j\pm}^{(1)}+\\
\sum\limits_{j\in R1,R2,R3}\Theta(t-t^{(1,2,3)}_j)\hat{f}_{j\pm}^{(1)}.
 \label{langevin lower state population
collective}
\end{eqnarray}

It should be pointed out that statistics of arriving times,
$t_j^{1,2,3}$ is defined by the character of the driving process and
might influence states of the emitted modes. This statistics
affects correlation properties of Langevin operators, which are
to be discussed in the next Subsection.

\subsection{Correlations functions for collective Langevin operators}

Notwithstanding the fact that  collective variables are composed
only from emitter operators of the same region, operators of
collective Langevin forces for different regions are not
independent and their cross-correlations are not always zero. It
occurs because of the presence of the same driving source for all
three regions. Naturally, such a driving partition can give rise
to cross-region correlations. Here we consider them generalizing a
simple and illustrative procedure described in details in
Ref.\cite{scully}. Using it, one obtains that
\begin{eqnarray}
\nonumber \left\langle \sum\limits_{j\in Rx,k\in Ry}\delta(t-t^{(x)}_j)\delta(t-t^{(y)}_k)\hat{n}_{j\pm}^{(2)}\hat{n}_{k\pm}^{(2)}\right\rangle_s=\\
(R_x+R_x^2)\delta_{xy}-\frac{R_xR_y}{R}\frac{p}{2},
\label{correlation langevin upper state population collective}
\end{eqnarray}
where the parameter $p$ describes the regularity of the pump and
$R=R_1+R_2+R_3$. The value $p=0$ corresponds to the Poissonian
pump (for example, as it is for pumping the active region with the
external laser field in Ref.\cite{pal2010}). The value $p=1$
corresponds to the regular pump when emitters arrive to the active
region with a constant rate. However, even in this case either "+"
or "-" subsystems are excited with equal probability; that is the
reason of having $p/2$ instead of $p$ in Eq.(\ref{correlation
langevin upper state population collective}). Values $0<p<1$
correspond to partially regular pumping. One can see from
Eq.(\ref{correlation langevin upper state population collective})
that for Poissonian pump noise correlations between region do not
arise, whereas for regular pumping one has cross-region
correlations. As we shall see below, such a partition noise
negates a noise-suppression effect of the regular pump. Photon
number squeezing in our two coupled VECSELs scheme arise solely
due to effect of the correlated loss, i.e. due to interference of
emission channels.

Thus, from Eqs.(\ref{langevin upper state population
collective}) it is possible to  get  second-order
correlation functions of the collective Langevin forces for
collective upper-state populations. For non-zero correlation functions of the same regions one has
\begin{eqnarray}
\label{correlation langevin upper state population collective1}
\nonumber
\left\langle\hat{F}_{2\pm}^{Y}(t)\hat{F}_{2\pm}^{Y}(\tau)\right\rangle=
\delta(t-\tau)R_{y}\left(1-\frac{R_{y}}{R}\frac{p}{2}\right)+\\
\nonumber
\delta(t-\tau)\left(\gamma_2Y_{2\pm}+\gamma_c(Y_{2\pm}+Y_{2\mp})\right), \\
\left\langle\hat{F}_{2\pm}^{Y}(t)\hat{F}_{2\mp}^{Y}(\tau)\right\rangle
=-\delta(t-\tau)\frac{R_{y}^2}{R}\frac{p}{2}-\\
\nonumber
\gamma_c\delta(t-\tau)\left(Y_{2\pm}+Y_{2\mp}\right),
\end{eqnarray}
where $Y=M,N,\Lambda$, and, simultaneously, $y=1,2,3$. For cross-correlation functions of different regions one has
\begin{eqnarray}
\nonumber
\left\langle\hat{F}_{2\pm}^{X}(t)\hat{F}_{2\pm}^{Y}(\tau)\right\rangle=
\left\langle\hat{F}_{2\pm}^{X}(t)\hat{F}_{2\mp}^{Y}(\tau)\right\rangle=\\
 -\delta(t-\tau)\left(\frac{R_xR_y}{R}\frac{p}{2}\right),
\quad X\neq Y, \quad x\neq y,
\label{correlation langevin upper state population collective11}
\end{eqnarray}
where $X,Y=M,N,\Lambda$, and, simultaneously, $x,y=1,2,3$. Variables without "hats" denote averages of corresponding
quantum variables, i.e. $\langle \hat{s}\rangle=s$, and averaging
is assumed to be done over the distribution of arrival times, too.

Since the emitters arrive completely excited, non-zero
second-order correlation functions of the collective Langevin
forces for collective lower-state populations are simpler than
Eqs. (\ref{correlation langevin upper state population
collective1},\ref{correlation langevin upper state population
collective11}):
\begin{eqnarray}
\left\langle\hat{F}_{1\pm}^{Y}(t)\hat{F}_{1\pm}^{Y}(\tau)\right\rangle=
\delta(t-\tau)\gamma_1{Y}_{1\pm},
 \label{correlation
langevin lower state population collective}
\end{eqnarray}
where $Y=M,N,\Lambda$.

For auto-correlation functions of collective polarizations one has
\begin{eqnarray}
\nonumber
\left\langle\left[\hat{F}_{\pm}^{Y}(t)\right]^{\dagger}\hat{F}_{\pm}^{Y}(\tau)\right\rangle=
\delta(t-\tau)\times \\
\left((2\gamma_{\bot}-\gamma_2-\gamma_c){Y}_{2\pm}+\gamma_c{Y}_{2\mp}+R_y\right), \\
\nonumber
\left\langle\hat{F}_{\pm}^{Y}(t)\left[\hat{F}_{\pm}^{Y}(\tau)\right]^{\dagger}\right\rangle=
\delta(t-\tau)(2\gamma_{\bot}-\gamma_1){Y}_{1\pm},
 \label{correlation langevin polarization
collective1}
\end{eqnarray}
where $Y=M,N,\Lambda$ and, simultaneously, $y=1,2,3$.

For non-zero cross-correlation functions of noises of collective polarization and
populations for the same region it
follows from Eqs.(\ref{langevin polarization
collective},\ref{langevin upper state population
collective},\ref{langevin lower state population collective})  that
\begin{eqnarray}
\nonumber
\left\langle\hat{F}_{\pm}^{X}(t)\hat{F}_{2\pm}^{Y}(\tau)\right\rangle=
\delta(t-\tau)(\gamma_2+\gamma_c){X}_{\pm}, \\
\left\langle\hat{F}_{\pm}^{X}(t)\hat{F}_{2\mp}^{Y}(\tau)\right\rangle=
-\delta(t-\tau)\gamma_c{X}_{\pm},\\
\nonumber
\left\langle\left[\hat{F}_{\pm}^{X}(t)\right]^{\dagger}\hat{F}_{2\mp}^{Y}(\tau)\right\rangle=
\delta(t-\tau)\gamma_1{X}_{\pm}^{\ast}.
 \label{correlation langevin polarization population
collective}
\end{eqnarray}
where $X=P,Q,\Xi$ and, simultaneously, $Y=M,N,\Lambda$.

Correlation functions derived in this Subsection will be used for
analyzing small fluctuations around stationary solutions of Eqs.
(\ref{field collective},\ref{polarization collective},\ref{upper
level population collective},\ref{lower level population
collective}).

\section{Quasiclassical equations and stationary solutions}

In this Section we analyze collective equations (\ref{field
collective},\ref{polarization collective},\ref{upper level
population collective},\ref{lower level population collective}) in
the quasiclassical limit, when one neglects quantum correlations
between variables, i.e. assumes that for any two variables
$\langle xy\rangle\approx \langle x\rangle\langle y\rangle$. We
shall consider the case where only two linearly and orthogonally
polarized modes persist in the whole system in the stationary
regime. Thus, we assume that for sufficiently large evolution times,
$t\rightarrow \infty$, only amplitudes
\[a_x=\frac{1}{\sqrt{2}}(a_++a_-)\]
and
\[b_y=\frac{i}{\sqrt{2}}(b_--b_+)\]
are non-zero. This regime was
realized in the experiment performed in Ref. \cite{pal2010}. It
should be noticed that reaching such a regime is not trivial,
because VECSEL modal dynamics can be quite involved. Generally,
one can have four elliptically polarized stationary modes in the
considered two coupled VECSEL system \cite{{kolobov2004},{modal
polarizations},{modal polarizations1}}. However, adjusting parameters of the scheme (in
particular, linear dichroism and birefringence) one can obtain the
desired regime with just two orthogonally polarized modes
surviving.

We assume also that the frequency difference between surviving
modes, $\Delta=\omega_a-\omega_b$, is very small on the scale set
by other parameters of the problem. So, we shall take that the
stationary regime is reached for times much less than
$\Delta^{-1}$.

Now let us demonstrate that in the quasiclassical limit and under
the time-scale hierarchy conditions (\ref{hierarchy}) collective
equations (\ref{field collective},\ref{polarization
collective},\ref{upper level population collective},\ref{lower
level population collective}) lead to standard quasiclassical
equations for modal intensities of the class-A laser. To start
with, let us re-write equations for upper-level populations
(\ref{upper level population collective}) in the quasiclassical
approximation
\begin{eqnarray}
\nonumber
\frac{d}{dt}M_{2\pm}=R_1-\gamma_2{M}_{2\pm}-\gamma_c({M}_{2\pm}-{M}_{2\mp})+\\
\nonumber
g_a({a}_{\pm}^{\ast}{P}_{\pm}+{P}^{\ast}_{\pm}{a}_{\pm}),\\
\label{upper level population classical}
\frac{d}{dt}{N}_{2\pm}=R_2-\gamma_2{N}_{2\pm}-\gamma_c({N}_{2\pm}-{N}_{2\mp})+\\
\nonumber
g_b({b}_{\pm}^{\ast}{Q}_{\pm}+{Q}^{\ast}_{\pm}{b}_{\pm}), \\
\nonumber
\frac{d}{dt}{\Lambda}_{2\pm}=R_3-\gamma_2{\Lambda}_{2\pm}-\gamma_c({\Lambda}_{2\pm}-{\Lambda}_{2\mp})-\\
\nonumber
((g_a{a}_{\pm}^{\ast}+g_b{b}_{\pm}^{\ast}){\Xi}_{\pm}+{\Xi}^{\ast}_{\pm}(g_a{a}_{\pm}+g_b{b}_{\pm})).
\end{eqnarray}

Due to the very rapid spin flips (according to the condition
(\ref{hierarchy})) in the stationary regime one has
${Y}_{2\pm}\approx{Y}_{2\mp}$, for $Y=M,N,\Lambda$. Also, if
one is not far from  threshold and the modal amplitudes are small,
it follows from Eqs. (\ref{upper level population classical}) that
in this regime the upper-level population in each region is  proportional to
the pumping rate in this region, i.e. ${Y}_{2\pm}\propto R_y$. Taking into
account Eq.(\ref{pump rates}) and the pump being common for all  regions of
the active media,  one comes to the following conclusion:
\begin{equation}
{\Lambda}_{2\pm}\approx
\sqrt{\xi_a}({M}_{2\pm}+{\Lambda}_{2\pm})\approx\sqrt{\xi_b}({N}_{2\pm}+{\Lambda}_{2\pm}).
\label{proportional populations}
\end{equation}

Now let us change the basis to the one rotating with the modal frequency (as
it was pointed above, we can safely take equal modal frequencies),
and introduce new collective variables corresponding to regions
coupled to certain modes,
\begin{eqnarray}
\nonumber {\mathcal P}_{\pm}=P_{\pm}+\Xi_{\pm}, \quad {\mathcal Q}_{\pm}=Q_{\pm}+\Xi_{\pm},\\
\nonumber
{\mathcal M}_{x\pm}=M_{x\pm}+\Lambda_{x\pm}, \quad {\mathcal N}_{x\pm}=N_{x\pm}+\Lambda_{x\pm},
\end{eqnarray}
where $x=1,2$. Then, our quasiclassical equations for modes are
\begin{eqnarray}
\nonumber \frac{d{a}_{\pm}}{dt}= - \kappa_a{a}_{\pm}
+(\kappa_{ap}+i\omega_{ap}) {a}_{\mp} + g_a{\mathcal P}_{\pm}, \\
\frac{d{b}_{\pm}}{dt}= - \kappa_b{b}_{\pm}
+(\kappa_{bp}+i\omega_{bp}){b}_{\mp} + g_b{\mathcal Q}_{\pm}.
\label{field collective classic}
\end{eqnarray}
For the collective polarization one has
\begin{eqnarray}
\nonumber\frac{d}{dt}{\mathcal P}_{\pm}=-(\gamma_{\bot}+i\nu){\mathcal P}_{\pm}+\\
\nonumber ({\mathcal M}_{2\pm}-{\mathcal M}_{1\pm})(g_a{a}_{\pm}+g_b\sqrt{\xi_a}b_{\pm}), \\
 \label{polarization collective classic}
\frac{d}{dt}{\mathcal Q}_{\pm}=-(\gamma_{\bot}+i\nu){\mathcal Q}_{\pm}+\\
\nonumber
({\mathcal N}_{2\pm}-{\mathcal N}_{1\pm})(g_b{b}_{\pm}+g_a\sqrt{\xi_b}{a}_{\pm}),
\end{eqnarray}
where $\nu=w-\omega_a$.

Since the lower state population is decaying very fast being on
the shortest time-scale in the considered system, near the
stationary regime one has ${X}_{2\pm}\gg {X}_{1\pm}$, for $X={\mathcal M},{\mathcal N}$.
Thus, the stationary values of collective polarizations can be
estimated from Eq.(\ref{polarization collective classic}) as
\begin{eqnarray}
\nonumber {\bar {\mathcal P}}_{\pm}\approx
\frac{{\mathcal M}_{2\pm}}{\gamma_{\bot}+i\nu}(g_a{a}_{\pm}+g_b\sqrt{\xi_a}b_{\pm}), \\
{\bar{\mathcal Q}}_{\pm}\approx
\frac{{\mathcal N}_{2\pm}}{\gamma_{\bot}+i\nu}(g_b{b}_{\pm}+g_a\sqrt{\xi_b}{a}_{\pm}).
 \label{polarization collective classic stationary}
\end{eqnarray}
From these equations it follows that proportionality relations
similar to (\ref{proportional populations}) hold for
polarizations too. It allows one to re-write the system
(\ref{upper level population classical}) as
\begin{eqnarray}
\nonumber
\frac{d}{dt}{\mathcal M}_{2\pm}=R_a-\gamma_2{\mathcal M}_{2\pm}-\gamma_c({\mathcal M}_{2\pm}-{\mathcal M}_{2\mp})+\\
\nonumber
\left((g_a{a}_{\pm}^{\ast}+g_b\sqrt{\xi_a}{b}_{\pm}^{\ast}){\mathcal P}_{\pm}+c.c\right),\\
\label{upper level population classical1}
\frac{d}{dt}{\mathcal N}_{2\pm}=R_b-\gamma_2{\mathcal N}_{2\pm}-\gamma_c({\mathcal N}_{2\pm}-{\mathcal N}_{2\mp})+\\
\nonumber
\left((g_b{b}_{\pm}^{\ast}+g_a\sqrt{\xi_b}{a}_{\pm}^{\ast}){\mathcal Q}_{\pm}+c.c.)\right),
\end{eqnarray}
where $R_a=R_1+R_3$, $R_b=R_2+R_3$. In the similar manner one can write equations for lower-level
populations, too.

Let us eliminate adiabatically low-level populations and polarizations,
taking into account the hierarchy of time-scales (\ref{hierarchy})
and assuming that the system is not far from the threshold. Then, from
Eqs.(\ref{field collective classic},\ref{polarization collective
classic stationary},\ref{upper level population classical1}) it is
easy to obtain the following equations for intensities of
surviving modes
\begin{eqnarray}
\nonumber \frac{d{I}_{a}}{dt}\approx -
2(\kappa_a-\kappa_{ap}){I}_{a} + \frac{r_{a}I_a}{d+c_a(I_a+\zeta_{ab}I_b)}, \\
\label{field collective classic1}
\frac{d{I}_{b}}{dt}\approx - 2(\kappa_b-\kappa_{bp}){I}_{b} +
\frac{r_{b}I_b}{d+c_b(I_b+\zeta_{ba}I_a)},
\end{eqnarray}
where $I_a=|a_x|^2$, $I_b=|b_y|^2$ and
\begin{eqnarray}
\nonumber
d=\gamma_2\left(1+\frac{\nu^2}{\gamma_{\bot}^2}\right),\quad r_x=\frac{2g_x^2}{\gamma_{\bot}}R_x, \\
\nonumber 
c_x=\frac{g_x^2}{\gamma_{\bot}}, \quad\zeta_{xy}=\xi_x\frac{g_y^2}{g_x^2},
\end{eqnarray}
where $x,y=a,b$ and $x\neq y$. Equations (\ref{field collective classic1}) were
derived under the condition that modal amplitudes are sufficiently
small, so that
\begin{equation}
d\gg c_x(I_x+\zeta_{xy}I_y),
\label{small field amplitudes condition}
\end{equation}
and quantities of the second and higher orders in $c_xI_x$ were
neglected.

Equations (\ref{field collective classic1}) are equations for modal
intensities of class-A lasers \cite{laser book}. They give the
following stationary solutions for modal intensities
\begin{eqnarray}
\nonumber \bar{I}_x=\left[g_x^2(1-\xi_a\xi_b)\right]^{-1}\times \\
\label{stationary modal intensities}
\left[\frac{g_x^2R_x}{k_x-k_{xp}}-\xi_y\frac{g_y^2R_y}{k_y-k_{yp}}-d\gamma_{\bot}(1-\xi_y)\right].
\end{eqnarray}
These stationary intensities are
non-zero when pumping rates exceed the threshold values
\begin{equation}
\bar{R}_x=\frac{d\gamma_{\bot}}{g_x^2}(\kappa_x-\kappa_{xp}).
\label{driving rates threshold values}
\end{equation}

Finally, let us write down the stationary values of upper-state
populations
\begin{eqnarray}
\nonumber
\bar{\mathcal M}_{2\pm}\approx
\frac{d}{\gamma_2}\frac{R_{a}}{d+c_a(I_a+\zeta_{ab}I_b)},\\
\label{stationary population}
\bar{\mathcal N}_{2\pm}\approx
\frac{d}{\gamma_2}\frac{R_{b}}{d+c_b(I_b+\zeta_{ba}I_a)}.
\end{eqnarray}

It is to be noticed that the threshold values of pumping rate
given by Eq.(\ref{driving rates threshold values}) are rather
large for the considered system being outside the "good cavity"
limit (this limit holds for $|g_x|>\kappa_x-\kappa_{xp}$). These
values are proportional to $\gamma_2$ and for a small detuning,
$|\nu|\ll\gamma_{\bot}$, they are also proportional to
$\gamma_{\bot}$.

\begin{figure}
\epsfig{ figure=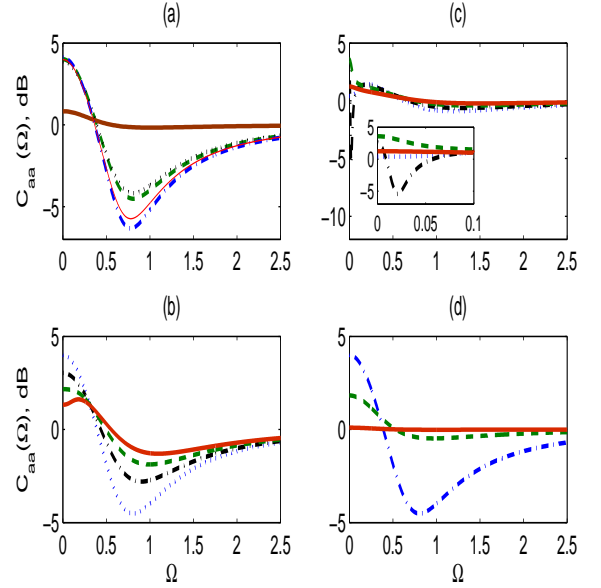,width=\linewidth,height=8cm} \caption
{(Color online) Spectra of the mode $a$ photocurrent, $C_{aa}(\Omega)$, for
values of system parameters  given by
Eq.(\ref{parameters}). (a) Spectra for different values of pumping
rate. Thick solid, dashed and dash-dotted lines correspond to
$R_a=R_b=1.001\bar{R}_a,1.01\bar{R}_a,1.011\bar{R}_a$ for the
coherent pumping, $p=0$;  dotted and thin solid lines
correspond to $R_a=R_b=1.01\bar{R}_a,1.011\bar{R}_a$ for the
regular pumping, $p=1$; the overlap is $\xi_a=\xi_b=0.8$. (b)
Spectra for larger overlaps of VECSELs active zones; solid, dashed, dash-dotted, dotted
lines correspond to $\xi_a=\xi_b=0.5,0.6,0.7,0.8$; the pumping is the coherent one,
$p=0$; $R_a=R_b=1.01\bar{R}_a$. (c) Spectra for smaller overlaps of VECSELs active zones;
solid, dashed, dash-dotted, dotted lines correspond to
 $\xi_a=\xi_b=0.1,0.2,0.3,0.4$, other parameters are as in the panel (b);
 the inset shows spectra for small frequencies for the same parameters as in the main panel.
(d) Spectra for different values of the
interaction constant;  dash-dotted, dashed and solid lines
correspond to
$g=0.01\kappa_a,0.05\kappa_a,0.1\kappa_a$; the
pumping is the coherent one, $p=0$; the overlap is
$\xi_a=\xi_b=0.8$; $R_a=R_b=1.01\bar{R}_a$. For all figures birefringence and the detuning
between modal frequencies and the emitter transition frequency are
taken to be zero, $\omega_{ap}=\omega_{bp}=0$, $\omega_{a}=w$;
$\Omega$ is given in units of $\kappa$. \label{fig2} }
\end{figure}

\begin{figure}
\epsfig{ figure=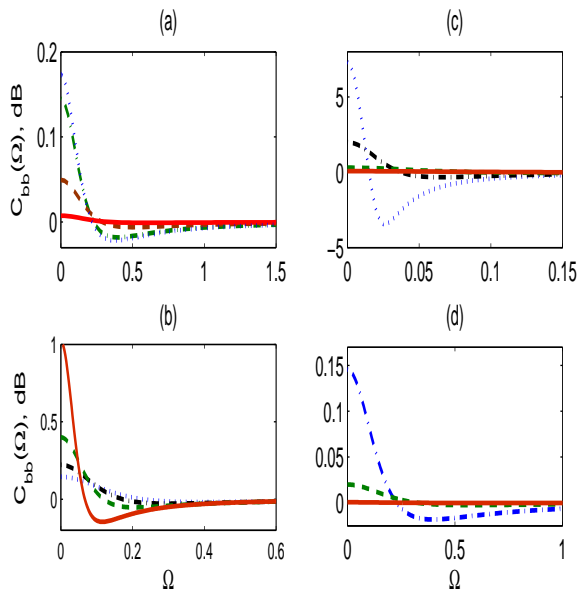,width=\linewidth,height=8cm} \caption
{(Color online) Spectra of the mode $b$ photocurrent, $C_{bb}(\Omega)$, for
values of system parameters  given by
Eq.(\ref{parameters}). (a) Spectra for different values of pumping
rate. Solid, dashed, dash-dotted and dotted lines correspond to
$R_a=R_b=1.001\bar{R}_a,1.005\bar{R}_a,1.01\bar{R}_a,1.011\bar{R}_a$;
the overlap is $\xi_a=\xi_b=0.8$, $p=0$ (the parameters are as in Fig.\ref{fig2}(a), coherent pumping). (b) Spectra for larger overlaps
of VECSELs active zones; solid, dashed, dash-dotted, dotted
lines correspond to $\xi_a=\xi_b=0.5,0.6,0.7,0.8$; other parameters are as in Fig.\ref{fig2}(b). (c) Spectra for
smaller overlaps of VECSELs active zones; solid, dashed, dash-dotted, dotted lines correspond to
 $\xi_a=\xi_b=0.1,0.2,0.3,0.4$, other parameters are as in the panel (b). (d) Spectra for different values
of the interaction constant; dash-dotted, dashed and solid lines
correspond to
$g=0.01\kappa_a,0.05\kappa_a,0.1\kappa_a$; other parameters are as in Fig.\ref{fig2}(d). For all figures the
pumping is the coherent one, $p=0$.  \label{fig3} }
\end{figure}

\section{Spectra of the quantum fluctuations}
\label{spectra}

So, let us assume that our system of two coupled VECSELs is close
to the stationary regime, and  consider small fluctuations
of the output modes. To this end we assume that each operator,
$\hat{x}$, representing a variable of the system can be written as
\[\hat{x}=\bar{x}+\delta x,
\]
where $\bar{x}$ denote the scalar stationary value and $\delta x$
is the operator describing quantum fluctuation and satisfying the
same commutation relation as the original operator $\hat{x}$. We
assume that $\langle\delta x\rangle=0$. Also, for simplicity sake
we shall assume that the stationary amplitudes of surviving field
modes are real in the basis rotating with the frequency $\omega_a$
(which can be safely assumed for such a situation
\cite{{kolobov2002},{kolobov2004}})).

After linearizing equations (\ref{field
collective},\ref{polarization collective},\ref{upper level
population collective},\ref{lower level population collective})
with respect to quantum fluctuation operators (similarly to as it
was done, for example, in Refs.
\cite{{kolobov2002},{kolobov2004}}), and eliminating adiabatically
fluctuations of the lower level populations, one obtains the
following system
\begin{eqnarray}
\frac{d}{dt}\overrightarrow{X}=\mathbf{D}\overrightarrow{X}+\overrightarrow{Z},
\label{system for small fluctuation time domain}
\end{eqnarray}
where elements of the vector $\overrightarrow{X}$ are quantum
fluctuation operators of system variables, and elements of the
vector $\overrightarrow{Z}$ are operators of corresponding
Langevin forces (see Appendix for the coefficients of these
vectors and the matrix $\mathbf{D}$, which is a $26\times26$
matrix).

Our aim is to investigate the spectrum of photocurrents produced by
modes going out of our VECSEL system (for example,in the scheme
outlined in Ref.\cite{pal2010}). For simplicity sake, let us
assume that our photodetectors are of unit efficiency, and losses
of surviving modes are caused solely by leaking through the
partially transparent mirror. So, the photocurrent is  directly
proportional to the number of photons of the corresponding
mode. Measuring spectra of individual current fluctuation or
cross-correlation fluctuation spectra, we are getting quantities
proportional to the following ones
\begin{eqnarray}
\langle[\delta I_{x}({\Omega})\delta I_{y}({\Omega})]\rangle=
\int\limits^{+\infty}_{-\infty}dte^{i\Omega t}\langle \delta
I_{x}(0)\delta I_{y}(t) \rangle, \label{photocurrent spectrum
general}
\end{eqnarray}
where $\langle \delta I_{x}(0)\delta I_{y}(t) \rangle$ is the
correlation function of the photocurrent fluctuation, $x=a,b$.
Operators $\delta I_x=I_x-\langle I_x\rangle$ describe
fluctuations of the photocurrent of output modes.

Implementing the standard input/output formalism for expressing
averages for modes outside the cavity through averages of modes
inside the cavity (see Refs.\cite{{colett},{kolobov1991}}),
one can obtain from Eq.(\ref{photocurrent spectrum general}) the
following expressions for the normalized photocurrent fluctuation
spectra
\begin{eqnarray}
\nonumber
 C_{xx}(\Omega)={\langle[\delta I_{x}({\Omega})\delta
I_{x}({\Omega})]\rangle}/{\langle I_x\rangle}=\\
1+4(\kappa_x-\kappa_{xp})d_{xx}(\Omega), \label{photocurrent
spectra}
\end{eqnarray}
and for the normalized cross-correlation function of photocurrrent
fluctuation of two output modes
\begin{eqnarray}
\nonumber
 C_{ab}(\Omega)=4\sqrt{(\kappa_a-\kappa_{ap})(\kappa_b-\kappa_{bp})}\times
 \\
 \frac{d_{ab}(\Omega)}{\sqrt{C_{aa}(\Omega)C_{bb}(\Omega)}},
\label{photocurrent cross correlation spectrum}
\end{eqnarray}
where quantities $d_{xy}$ are defined through second-order
correlation functions of modal operators as
\begin{eqnarray}
\langle \hat{d}_{x}(\Omega)\hat{d}_{y}(\Omega')
\rangle=d_{xy}(\Omega)\delta(\Omega+\Omega') \label{fourier
correlations},
\end{eqnarray}
with operators
\begin{eqnarray}
\nonumber \hat{d}_{a}(\Omega)=\frac{1}{2}(\delta
a_+(\Omega)+\delta a_-(\Omega)+h.c.),\\
\hat{d}_{b}(\Omega)=\frac{1}{2}(\delta b_-(\Omega)-\delta
b_+(\Omega)+h.c.).
 \label{input operators}
\end{eqnarray}
In Eq.(\ref{input operators}) operators $\delta x_+(\Omega)$ are elements of the
Fourier-transformed solution of the system (\ref{system for small
fluctuation time domain}):
\begin{eqnarray}
\overrightarrow{X}(\Omega)=\left[\mathbf{D}-\Omega\right]^{-1}\overrightarrow{Z}(\Omega),
\label{system for small fluctuation frequency domain}
\end{eqnarray}
where
\[\overrightarrow{Z}(\Omega)=\int\limits^{+\infty}_{-\infty}dte^{i\Omega t}\overrightarrow{Z}(t).\]

\begin{figure}
\epsfig{ figure=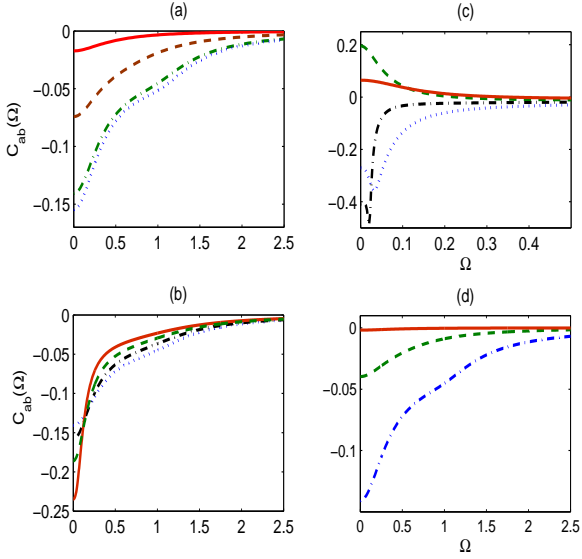,width=\linewidth,height=8cm}
\caption{(Color online) Cross-correlation function, $C_{ab}(\Omega)$, for values
of system parameters close to ones given by Eq.(\ref{parameters}).
(a) Cross-correlation function for different values of pumping
rate. Solid, dashed, dash-dotted and dotted lines correspond to
$R_a=R_b=1.001\bar{R}_a,1.005\bar{R}_a,1.01\bar{R}_a,1.011\bar{R}_a$;
other parameters are as in Fig.\ref{fig3}(a). (b) Cross-correlation
function for larger overlaps of VECSELs active zones; solid, dashed, dash-dotted, dotted
lines correspond to $\xi_a=\xi_b=0.5,0.6,0.7,0.8$; other parameters are as in Fig.\ref{fig3}(b) (c) Cross-correlation function for
smaller overlaps of VECSELs active zones; solid, dashed, dash-dotted, dotted lines correspond to
 $\xi_a=\xi_b=0.1,0.2,0.3,0.4$, other parameters are as in the panel (b). (d) Cross-correlation function for
different values of the interaction constant; solid, dashed and dash-dotted lines
correspond to
$g=0.01\kappa_a,0.05\kappa_a,0.1\kappa_a$; other parameters are as in Fig.\ref{fig2}(d).
 \label{fig4} }
\end{figure}

For illustration of results described by Eqs. (\ref{photocurrent
spectra}-\ref{system for small fluctuation frequency domain}), let
us consider the simplest symmetrical situation, when both
coupling, overlaps and modal decay rates are equal: $g_a=g_b\equiv
g$, $\xi_a=\xi_b\equiv\xi$, $\kappa_a=\kappa_b\equiv\kappa$,
$\kappa_{ap}=\kappa_{ap}=\kappa_{p}$; and assume the following
realistic hierarchy of time-scales:
\begin{eqnarray}
\nonumber g=0.1\kappa, \quad \kappa_p=0.5\kappa, \quad
\gamma_2=10\kappa, \\
\gamma_c=10^2\kappa, \quad \gamma_{\bot}=10^3\kappa.
\label{parameters}
\end{eqnarray}
Results of numerical simulation with parameters close to the ones
given by Eq.(\ref{parameters}) are given in Fig.\ref{fig2} for the
normalized spectrum of the mode $a$, in Fig.\ref{fig3} for the
normalized spectrum of the mode $b$, and in Fig.\ref{fig4} for
the cross-correlation function (\ref{photocurrent cross
correlation spectrum}).

First of all, notice that spectra of photocurrent fluctuations of
output modes $a$ and $b$ are drastically different
(notwithstanding the fact that for the chosen symmetric case the
stationary values of modal intensities are the same). An explanation for this
phenomenon can be guessed even from the toy model described in
Subsection \ref{simple model}: from Eq. (\ref{model master
equation3}) one can see that the symmetric modal superposition is
subjected to nonlinearity, whereas the antisymmetric is not. So,
for orthogonally polarized modes one should expect different
noises. Then, it can be seen that spectral properties of
fluctuations are strongly dependent on the overlap between active
regions of VECSELs. It is possible to distinguish three different regimes of
fluctuations in dependence on the overlap: the strong
(approximately $0.5\leq\xi<1$), intermediate ($0.1\leq\xi\leq
0.5$) and weak ($\xi\leq0.1$) overlap.

Let us start with the discussion of noise features common for all
three regimes. In Fig.\ref{fig2}(a) it is shown how the photocurrent noise
of the output mode $a$ for fixed overlap is
suppressed more with increase of the pumping rate. Quite significant suppression of the photocurrent
noise can be reached (say, about $90\%$). However, with the used values of
parameters (chosen to make the system operating as a class-A
laser) one quickly comes out of the region of applicability of the
approximation used to derive the system (\ref{system for small
fluctuation time domain}). This feature was commented upon in the
previous Section. So, for the pumping rate exceeding the threshold
value (\ref{driving rates threshold values}) only by $1.5\%$ the
condition (\ref{small field amplitudes condition}) is already
breaking down. It is remarkable that regularity of the pump can
even diminish noise reduction for the case (which is also illustrated
in Fig. \ref{fig2}(a)). The reason for this is simple: the common pump for
both coupled VECSEL systems induces rather strong partition noise.
It obliterates the effect of regularity. Whereas for the coherent
Poissonian pumping partition noise is not present. Also, it is
worth noting that the harmful influence of regularity is more
pronounced with increase of the pumping rate, since the terms
describing the partition noise grow linearly with the pumping rate
(see Eq.(\ref{correlation langevin upper state population
collective1}).

The photocurrent fluctuation spectrum for the mode $b$ depends on
the pumping rate in the same way as it is for the mode $a$ for the
fixed value of the overlap (Fig.\ref{fig2}(b)).

Another feature common for all three regimes is dependence of the
exhibited noise suppression on respective modal losses. It should
be stressed out that for any significant suppression of the
photocurrent noise to appear one needs having rather good cavity
for their VECSELs. When one goes far away from this limit,
noise reduction degrades significantly (Fig.\ref{fig2}(d) and
Fig.\ref{fig3}(d)). For the coupling constant, $g$, just two
orders of magnitude lower than the modal loss rate,
$\kappa-\kappa_{p}$, any noise suppression is already absent. This
result is rather expected, since strong modal loss is bound to
destroy quickly interference effects leading to non-classicality
(which is well-known fact for those trying to produce
non-classical states with Kerr-type nonlinearity
\cite{{yurke},{glancy}}).

Now let us discuss the dependence of photocurrent noise spectra on
the overlap between regions of the active media. Fig.\ref{fig2}(c)
and Fig.\ref{fig3}(c) show that the mechanism of photocurrent
noise suppression for the case is indeed the correlated loss. In
the regime of weak overlap (taking place approximately for
$\xi\leq 0.1$) for both output modes suppression is absent. In
this regime the system behaves as two nearly uncoupled VECSELs
with coherent pumping and exhibits weak bunched noise diminishing
with increase of the frequency (which is typical for the
individual single-mode class-A VECSELs considered here
\cite{{a-class1},{a-class2}}).

In the regime of the strong overlap (approximately
$0.5\leq\xi<1$), the output mode $a$ shows much larger noise suppression
than the mode $b$ (which is a manifestation of
noise dependence on the modal polarization mentioned earlier).
When the overlap becomes smaller, noise for the mode $a$ and
$b$ behaves in the opposite way. Noise decreases for the mode $a$ and
increases for the mode $b$ (see Fig.\ref{fig2}(b) and
Fig.\ref{fig3}(b)). An increase of noise suppression for the mode $b$
continues well into the intermediate regime, $0.1\leq\xi\leq 0.5$,
reaching the maximum when approximately half of the active media
regions overlap. Whereas in this regime noise reduction for the mode
$a$ is rather weak.

Since our non-classical noise suppression effect is produced by
the interference of the emission channels of emitters arriving to
the active media, i.e. by coupling of modes to the same emitters
subjected to strong losses,  it is naturally to expect a strong
intermodal correlation arising as the result of this process (as
it is pointed out by the simple model considered in Subsection
\ref{simple model}, where it is demonstrated how the coupling between
the modes arises for this case). Due to different behavior of the
photocurrent fluctuation spectra for modes $a$ and $b$ one expects
having rather non-trivial cross-correlation spectrum
(\ref{photocurrent cross correlation spectrum}). Having nearly
Poissonian noise of the mode $b$ and strongly suppressed photocurrent
noise of the mode $a$ in the larger overlap regime, it is quite intuitive to see
anticorrelations between modes  in this regime
(Fig.\ref{fig4}(a,b)). Also, it is not surprising to have these
anticorrelations diminishing with decrease of the pumping rate or
emitter-field interaction constants for the fixed overlap (see
Fig.\ref{fig4}(a) and (d)).

However, notice that the maximum anticorrelation for the large
overlap is reached for small frequency regions where the
photocurrent fluctuations spectra for both modes  are
actually super-Poissonian (Fig.\ref{fig4}(a)). This is also a
manifestation of the correlated loss as an interference of the
emission channels for emitters arriving to the active medium. An
individual emitter emits a photon either to one mode or another,
and due to a large population and polarization losses the
probability to have the photon re-absorbed by the emitter is very
low. This process might not prevent noises of individual modes from exhibiting
bunching, but it induces anticorrelations between modes.

Maximal anticorrelation is reached in the intermediate regime where
the photocurrent spectrum of the mode $b$ exhibit maximal
noise reduction. With further decrease of the overlap
anticorrelations are eventually washed out tending to small positive
correlations in the small overlap regime (Fig.\ref{fig4}(c)). Notice that maximal
increase of the anticorrelations has obviously resonant character depending strongly on
the overlap and the frequency; the maximum anticorrelation is reached for the overlap corresponding
approximately to all three regions $R_j$ being equally sized (see Fig.\ref{fig1}).

It should be noticed that correlations between modes are even more
sensitive to modal losses than non-classicality of individual
modes (see Fig.\ref{fig4}(d)). As it should be expected, modal
losses into independent additional dissipative reservoir are prone
to obliterate quickly the effect of the correlated loss (i.e.
coupling to the same reservoir) \cite{mogilevtsev kerr 2009}. One
should really be not far from the good cavity limit to have
significant correlations between modes in two coupled VECSELs
considered here.

\section{Conclusions}

In this work we have demonstrated that the system of two coupled
VECSELs can indeed be a class-A laser, if it is sufficiently close
to the threshold. We have developed a simple model based on the
quantum Langevin equations that leads to the semiclassical
equations for the intensities of two surviving orthogonally
linearly polarized output modes similar to the equations
describing the class-A laser. We have demonstrated that both
suppression of photocurrent fluctuation noise below the standard quantum limit  and strongly
negative cross-correlation (anticorrelation) are possible in the
system. It is remarkable that the mechanism of noise suppression
arising in this case is quite different from the already
well-known mechanism of inducing noise suppression through the
regular pumping. In our case  non-classicality arises through
correlated loss, i.e. due to simultaneous coupling of both modes
to the same emitter and quick decay of populations and
polarization of this emitter. It is demonstrated by disappearance
of non-classicality for small overlap between active regions of
VECSELs, when the dynamics of each VESCEL is essentially as for an
individual independent laser. For both strong and moderate overlap
it is possible to reach quite strong suppression of photocurrent
fluctuations (5-10 dB) and large anti-correlations. Notice, that
regularity of the pumping does not enhance non-classicality
available with coherent pumping. Actually, as a consequence of
partition noise arising due to the pump being common for both
VECSELs, regularity of the pumping can actually be harmful for
non-classicality.

\acknowledgments

The authors (D.~M. and D.~H.) are very thankful to the Laboratoire
PhLAM, Universit\'{e} Lille 1 for warm hospitality. D.~M. and D.~H.
acknowledge financial support from CNRS, BRFFI and the research program "Convergence" of NASB.  We also acknowledge financial support from RFFI, (grants 12-02-00181a and 13-02-00254a) (Yu. M. G.), the programm ERA.Net RUS (project NANOQUINT),  the grant of the St. Petersburg State University, 11.38.70.2012, and the FAPESP grant  2014/21188-0 (D.M.).  The authors
thank N. Larionov for fruitful discussions. 

\clearpage
\appendix
\section{Linearized equations for quantum fluctuations}

Here we write down vectors of variables and non-homogeneous part
of the linearized system (\ref{system for small fluctuation time
domain}):
\begin{eqnarray}
\overrightarrow{X}=
\begin{pmatrix}
 \delta a_+ \\
  \delta a_-  \\
 \delta b_+  \\
  \delta b_-  \\
  \delta a_+^{\dagger} \\
  \delta a_-^{\dagger}  \\
  \delta b_+^{\dagger}  \\
  \delta b_-^{\dagger}\\
\delta P_{+} \\
  \delta P_{-}  \\
 \delta Q_{+} \\
  \delta Q_{-} \\
   \delta \Xi_{+} \\
  \delta \Xi_{-} \\
   \delta P_{+}^{\dagger} \\
  \delta P_{-}^{\dagger}\\
\delta Q_{+}^{\dagger} \\
  \delta Q_{-}^{\dagger}\\
\delta \Xi_{+}^{\dagger} \\
  \delta \Xi_{-}^{\dagger}\\
  \delta M_{2+}\\
  \delta M_{2-}\\
  \delta N_{2+}\\
  \delta N_{2-}\\
  \delta \Lambda_{2+}\\
  \delta \Lambda_{2-}\\
\end{pmatrix}, \quad
\overrightarrow{Z}=
\begin{pmatrix}
  f_{a+} \\
  f_{a-}  \\
 f_{b+} \\
  f_{b-}  \\
  f_{a+}^{\dagger} \\
  f_{a-}^{\dagger}  \\
 f_{b+}^{\dagger} \\
  f_{b-}^{\dagger}  \\
F^P_{+} \\
 F^P_{-}  \\
 F^Q_{+} \\
 F^Q_{-}  \\
   F^{\Xi}_{+} \\
 F^{\Xi}_{-}  \\
   [F^P_{+}]^{\dagger} \\
 [F^P_{-}]^{\dagger}  \\
 [F^Q_{+}]^{\dagger} \\
 [F^Q_{-}]^{\dagger}  \\
   [F^{\Xi}_{+}]^{\dagger} \\
 [F^{\Xi}_{-}]^{\dagger}  \\
  F^{M}_{2+}\\
  F^{M}_{2-}\\
  F^{N}_{2+}\\
  F^{N}_{2-}\\
  F^{\Lambda}_{2+}\\
  F^{\Lambda}_{2-}\\
\end{pmatrix}
. \label{vectors}
\end{eqnarray}

The $26\times26$ matrix $\mathbf{D}$ can be represented in the
block form as
\begin{eqnarray}
\mathbf{D}=
\begin{pmatrix}
  D_{AA} & D_{AP} & D_{AN} \\
  D_{PA} & D_{PP} & D_{PN} \\
  D_{NA} & D_{NP} & D_{NN}
\end{pmatrix}
. \label{matrix D}
\end{eqnarray}

Non-zero elements of $8\times8$ matrix $D_{AA}$ are
\begin{eqnarray}
\nonumber D_{AA}(j,j)=-\kappa_a \quad \mathrm{for}\quad 1\leq j\leq 4, \\
\nonumber
D_{AA}(1,2)=D_{AA}(5,6)^{\ast}=\kappa_{ap}+i\omega_{ap},\\
\nonumber
D_{AA}(j,j)=-\kappa_b \quad \mathrm{for}\quad 5\leq j\leq 8, \\
\nonumber D_{AA}(3,4)=D_{AA}(7,8)^{\ast}=\kappa_{bp}+i\omega_{bp},
 \label{daa}
\end{eqnarray}
and $D_{AA}(j,k)=D_{AA}(k,j)$. The elements $8\times6$ matrix
$D_{AN}$ are zeros. Non-zero elements of the $8\times12$ matrix
$D_{AP}$ are

\begin{eqnarray}
\nonumber D_{AP}(j,j)=D_{AP}(j,j+5)=g_a,\quad j=1,2,5,6 \\
\nonumber D_{AP}(j,j)=D_{AP}(j,j+2)=g_b,\quad j=3,4,7,8.
 \label{dap}
\end{eqnarray}
Non-zero elements of the $12\times8$ matrix $D_{PA}$ are

\begin{eqnarray}
\nonumber D_{PA}(j,j)=D_{PA}(j+6,j+4)=g_a\bar{M}_{2\pm},\quad j=1,2; \\
\nonumber
D_{PA}(j,j)=D_{PA}(j+6,j+4)=g_b\bar{N}_{2\pm}, \quad j=3,4; \\
\nonumber
D_{PA}(j+4,j)=D_{PA}(j+10,j+4)=g_a\bar{\Lambda}_{2\pm}, \quad j=1,2; \\
\nonumber
 D_{PA}(j+2,j)=D_{PA}(j+8,j+3)=g_b\bar{\Lambda}_{2\pm},\quad j=3,4;
 \label{dpa}
\end{eqnarray}
where the stationary values of upper-level populations
are given by Eq.(\ref{stationary
population}). Non-zero elements of the $12\times12$ matrix
$D_{PP}$ are diagonal:
$D_{PP}(j,j)=[D_{PP}(j+4,j+4)]^{\ast}=-\gamma_{\bot}-i\nu$ for
$1\leq j\leq4$. Non-zero elements of the $12\times6$ matrix
$D_{PN}$ are

\begin{eqnarray}
\nonumber D_{PN}(1,1)=[D_{PN}(7,1)]^{\ast}=g_a\bar{a}_+,\\
\nonumber
D_{PN}(2,2)=[D_{PN}(8,2)]^{\ast}=g_a\bar{a}_-, \\
\nonumber
D_{PN}(3,3)=[D_{PN}(9,3)]^{\ast}=g_b\bar{b}_+, \\
\nonumber
D_{PN}(4,4)=[D_{PN}(10,4)]^{\ast}=g_b\bar{b}_- ,\\
\nonumber
D_{PN}(5,5)=[D_{PN}(11,5)]^{\ast}=g_a\bar{a}_++g_b\bar{b}_+ , \\
\nonumber
D_{PN}(6,6)=[D_{PN}(12,6)]^{\ast}=g_a\bar{a}_-+g_b\bar{b}_- ,
 \label{dpn}
\end{eqnarray}
where stationary modal amplitudes are taken to be real and
positive (as it was pointed out in Section \ref{spectra}), and
given by Eq.(\ref{stationary modal intensities}).

Non-zero elements of the $6\times8$ matrix $D_{NA}$ are
\begin{eqnarray}
\nonumber
[D_{NA}(1,1)]^{\ast}=D_{NA}(1,5)=g_a\bar{P}_{+}, \\
\nonumber
[D_{NA}(2,2)]^{\ast}=D_{NA}(2,6)=g_a\bar{P}_{-}, \\
\nonumber
[D_{NA}(3,3)]^{\ast}=D_{NA}(3,7)=g_b\bar{Q}_{+}, \\
\nonumber
[D_{NA}(4,4)]^{\ast}=D_{NA}(4,8)=g_b\bar{Q}_{-}, \\
\nonumber
[D_{NA}(5,1)]^{\ast}=D_{NA}(5,5)=g_a\bar{\Xi}_{+}, \\
\nonumber
[D_{NA}(6,3)]^{\ast}=D_{NA}(6,7)=g_a\bar{\Xi}_{-}, \\
\nonumber
[D_{NA}(5,2)]^{\ast}=D_{NA}(5,6)=g_b\bar{\Xi}_{+}, \\
\nonumber [D_{NA}(6,4)]^{\ast}=D_{NA}(6,8)=g_b\bar{\Xi}_{-},
\end{eqnarray}
where stationary polarizations are given by Eqs.(\ref{polarization
collective classic stationary}). Non-zero elements of the
$6\times12$ matrix $D_{NP}$ are

\begin{eqnarray}
\nonumber
[D_{NP}(1,1)]^{\ast}=D_{NP}(1,7)=g_a\bar{a}_{+}, \\
\nonumber
[D_{NP}(2,2)]^{\ast}=D_{NP}(2,8)=g_a\bar{a}_{-}, \\
\nonumber
[D_{NP}(3,3)]^{\ast}=D_{NP}(3,9)=g_b\bar{b}_{+} \\
\nonumber
[D_{NP}(4,4)]^{\ast}=D_{NP}(4,10)=g_b\bar{b}_{-}, \\
\nonumber
[D_{NP}(5,5)]^{\ast}=D_{NP}(5,11)=g_a\bar{a}_{+}+g_b\bar{b}_{+}, \\
\nonumber
[D_{NP}(6,6)]^{\ast}=D_{NP}(6,12)=g_a\bar{a}_{-}+g_b\bar{b}_{-}.
\end{eqnarray}

Finally, non-zero elements of the $6\times6$ matrix $D_{NN}$ are
$D_{NN}(j,j)=-\gamma_2-\gamma_c$ for $j=1,2\ldots6$;
$D_{NN}(j,j+1)=\gamma_c$ for $j=1,3,5$; also
$D_{NN}(j,k)=D_{NN}(k,j)$ for all $j,k$.

\end{document}